%Paper: hep-th/9309044
%From: poly@dxcern.cern.ch (Alexios Polychronakos)
%Date: Wed, 8 Sep 1993 14:08:15 +0200

%%%%%%%%%%%%%%%%%%%%%%%%%%%%%%%%%%%%%%%%%%%%%%%%%%%%%%%%%%%%%%%%%%%%%%%%%%%%%%%
%%%%% This Tex file inputs phyzzx.       				  %%%%%
%%%%%%%%%%%%%%%%%%%%%%%%%%%%%%%%%%%%%%%%%%%%%%%%%%%%%%%%%%%%%%%%%%%%%%%%%%%%%%%
\input phyzzx
\def\Om{\Omega}

\def\CRp{C_{2R'}}

\def\th{\theta}

\def\izl{\int_0^L dx}
\def\Ad{\dot A_1}
\def\Wd{\dot W}

\def\de{\delta}
\def\eps{\epsilon}
\def\bpsi{\bar\psi}
\def\Lam{\Lambda}
\def\Udag{U^\dagger}
\def\dU{\dot U}
\def\UddU{\Udag\dU}
\def\Lam{\Lambda}
\def\Lamdag{\Lam^\dagger}

\def\QCD{${\rm QCD}_2$}

\def\dag{\dagger}

\def\adI{a^{\dagger I}}
\def\adJ{a^{\dagger J}}
\def\bdI{b^{\dagger I}}

\Pubnum={CERN-TH-6994/93\cr
UVA-HET-93-11\cr
hepth@xxx/9309044}
%\date={September 1993}
\titlepage
\title{Interacting Fermion Systems from Two Dimensional QCD}
\bigskip
\author {
Joseph~A.~Minahan\footnote\star
{Address after Sept. 1, 1993: Department of Physics, University of
Southern California, Los Angeles, CA 90089-0484, USA}}
\address{Department of Physics,  Jesse Beams Laboratory,\break
University of Virginia, Charlottesville, VA 22901, USA}
\andauthor{
Alexios P. Polychronakos\footnote\dagger
{poly@dxcern.cern.ch}}
\address{Theory Division, CERN\break
CH-1211, Geneva 23, Switzerland}
\bigskip
\abstract{
We consider two dimensional $U(N)$ QCD on the cylinder with a timelike
Wilson line in an arbitrary representation. We show that the theory is
equivalent to $N$ fermions with internal degrees of freedom which interact
among themselves with a generalized Sutherland-type interaction. By evaluating
the expectation value of the Wilson line in the original theory we explicitly
find the spectrum and degeneracies of these particle systems.
}
%\submit{}
\vfill
\endpage

\def\NP{{\it Nucl. Phys.\ }}
\def\PL{{\it Phys. Lett.\ }}
\def\PRD{{\it Phys. Rev. D\ }}
\def\PRB{{\it Phys. Rev. B\ }}
\def\PRA{{\it Phys. Rev. A\ }}
\def\PRL{{\it Phys. Rev. Lett.\ }}

\def\MPL{{\it Mod. Phys. Lett. A\ }}

\def\ZETF{{\it Zh. Eksp. Teor. Fiz.}}

\REF\Gross{D.~Gross, \NP {\bf B400} (1993) 161; LBL 33232, PUPT 1355,
hepth/9212148.}
\REF\JM{J.~Minahan, \PRD {\bf 47} (1993) 3430.}
\REF\GrTay{D.~Gross and W.~Taylor, \NP {\bf B400} (1993) 181.}
\REF\GrTayII{D.~Gross and W.~Taylor, \NP {\bf B403} (1993) 395.}
\REF\MP{J.~Minahan and A.~P.~Polychronakos, \PL {\bf B312} (1993) 155.}
\REF\Mike{M.~Douglas, RU-93-13, hep-th/9303159.}
\REF\Sut{B.~Sutherland, \PRA {\bf 4}, 2019 (1971); \PRA {\bf5}, 1372 (1972);
\PRL {\bf34}, 1083 (1975).}
\REF\APg{A.~P.~Polychronakos, \PL {\bf B266} (1991) 29.}
\REF\GN{A.~Gorsky and N.~Nekrasov, ITEP-20/93, hepth/9304047.}
\REF\Migd{A.~Migdal, \ZETF {\bf 69} (1975) 810.}
\REF\Rus{B.~Rusakov, \MPL {\bf5} (1990) 693.}
\REF\Kaw{N.~Kawakami, \PRB {\bf46} (1992) 1005; \PRB {\bf46} (1992) 3191.}
\REF\HaHal{Z.~N.~C.~Ha and F.~D.~M.~Haldane, Princeton preprint, 1992.}
\REF\MPII{J.~Minahan and A.~P.~Polychronakos, \PL {\bf B302} (1993) 265.}

There has been recent interest in two dimensional QCD, mainly in understanding
its string like properties [\Gross-\GrTayII].
One of the salient features of this theory
is that pure gauge $U(N)$ theory is equivalent to $N$ free fermions on a
circle [\MP,\Mike]. In this letter we consider the modification to this theory
when an immobile color source is added to the theory.  As far as the
path integral goes, this is equivalent to inserting a timelike Wilson
line in the theory.  We will show that the effect of this source is
to turn this into a theory of interacting fermions with internal degrees
of freedom that transform under a representation of some group $SU(n)$,
where $n$ is determined by the representation of the Wilson line. If
its Young tableau consists of $n$ rows then one can choose the
representations of the internal quantum numbers to be symmetric representations
of $SU(n)$. One can also choose the internal quantum numbers to transform
under antisymmetric representations of $SU(m)$, where $m$ is the number of
columns in the Young tableau. We will then derive the spectrum of this
interacting fermion theory, with its degeneracies, by comparing with the
expectation value of the Wilson loop in the original theory.

Consider $U(N)$ \QCD\ on the cylinder.
Let $L$ be the circumference of the cylinder and $T$ be its length in the time
direction.  Suppose that there is a Wilson line along the time direction with
spatial position $x=0$ in the representation $R$ of the group $U(N)$. Hence
this theory is described by having an immobile color source at position $x=0$.

Let us then consider the action for the continuum theory
$${1\over4}\int d^2x \tr F_{\mu\nu}F^{\mu\nu}+
\int dt \bpsi\Bigl(i\partial_t-gA^a_0(x=0)T^a_R +M\Bigr)\psi\eqn\action$$
where the $T^a_R$ are the generators of the group in representation
$R$, $\psi$ ($\bpsi$) is the annihilation (creation) operator
for the heavy color source and $M$ is its mass.

The euclidean path integral over the heavy fermion can be explicitly evaluated.
It can be easily obtained, however, by noticing that in the proper fermion
basis the theory consists of $d_R$ independent fermions with energies
$M-i{\omega_n \over T}$, where $\exp(i\omega_n )$ are the eigenvalues of the
timelike Wilson loop in the representation $R$ (notice that the gauge term
remains imaginary even in euclidean time). Integrating out $\psi$, then, will
give the partition function for the fermions which is
$$Z = \prod_{n=1}^{d_R} \left( 1 + e^{-TM+i\omega_n} \right) =
1 + e^{-TM} \sum_{n=1}^{d_R} e^{i\omega_n} + {\cal O}\Bigl( e^{-2TM}\Bigr)
\eqn\Z$$
To lowest order we have pure \QCD, while to order $e^{-TM}$ we obtain
the trace of the Wilson loop. Isolating the ${\cal O}(\exp(-TM))$ term in
the full partition function will then give us the theory with an
insertion of a timelike Wilson loop in representation $R$.

In the canonical formulation, the Hamiltonian in the gauge $A_0=0$ is
$$H=\half\izl\tr F_{01}^2 +M\bpsi\psi=\half\izl\tr \Ad^2 +M\bpsi\psi
\eqn\Hamgauge$$
with the overdot denoting a time derivative.
The $A_0$ equation of motion is now the constraint
$$D_1F_{10}=\partial_1\Ad+ig[A_1,\Ad]=gK\de(x-L+\eps),\eqn\gconstraint$$
where
$$K=\sum_{a=1}^{N^2 -1}K^a \tau^a ~,~~~ K^a =\bpsi T^a_R \psi\eqn\K$$
and $\tau^a$ are the $SU(N)$ generators in the fundamental representation.
For later convenience, we have actually moved the source over to the left by
a small amount $\eps$.

We can now proceed in a manner similar to that in [\MP].
Define a new variable $V(x)$,
$$V(x)=W_0^x\Ad(x)W_x^L,\eqn\Vdef$$
where
$$W_a^b={\rm P}e^{ig\int_a^bdxA_1}.\eqn\Wdef$$
Then \gconstraint\ can be written as
$$\partial_1V(x)=gW_0^{L-\eps}KW_{L-\eps}^L\de(x-L+\eps).\eqn\Veq$$
Thus $V(x)$ is constant until it reaches $x=L-\eps$, at which point
it jumps by $gW_0^{L-\eps}KW_{L-\eps}^L$, which in the limit $\eps\to0$
becomes $gWK$, where $W=W_0^L$.  Hence $V(0)=V(L)+gWK$, which implies that
$$[W,\Ad(0)]=gWK,\eqn\WAcomm$$
where we have used the periodicity of $A_1$ in $x$.

{}From the definitions \Vdef\ and \Wdef, we find the relation
$$\Wd=ig \izl W_0^x\Ad(x)W_x^L=ig \izl V(x),\eqn\Wdeq$$
and therefore using \Veq\ and \WAcomm\ and letting $\eps\to0$, one finds
$$\Wd=igL\Ad(0)W.\eqn\WdeqII$$
Equations \WAcomm\ and \WdeqII\ then imply that
$$[W,\Wd]=ig^2LK.\eqn\WWdeq$$
Because $V(x)=V(0)$ for $0\le x<L-\eps$, $\Ad(x)$ satisfies
$$\Ad(x)=W_0^x\Ad(0)W_x^0.\eqn\Adeq$$
Thus, using this relation along with \WdeqII, we can rewrite the Hamiltonian
in \Hamgauge\ as
$$H=-{1\over 2g^2L}\tr(W^{-1}\Wd)^2 +M\bpsi\psi.\eqn\Hammm$$

The operators $K^a$ provide a realization of the $SU(N)$ algebra in terms of
fermionic oscillators, which decomposes into irreducible representations. The
lowest one is the singlet (corresponding to the fermionic vacuum). For this
irrep the fermion mass term in \Hammm\ (which is a casimir of $SU(N)$)
vanishes. If the gauge group is $U(N)$, with the $U(1)$ coupling
given by $g/N$, then \Hammm\ becomes
the Hamiltonian for the one-dimensional unitary matrix model. The right-hand
side of \WWdeq\ also vanishes. We thus recover the one-dimensional matrix
model and constraint equivalent to \QCD\ [\MP]. This corresponds to the
${\cal O}(1)$ term in \Z. The next highest irrep contained in $K$ is $R$.
For this irrep the fermion mass term equals $M$ and thus it corresponds to the
order ${\cal O}(\exp(-TM))$ term in \Z. This is, then, the part corresponding
to the Wilson loop insertion, and in this case \WWdeq\ carries the
representation $R$.

We note here that we could have quantized $\psi$ as a bosonic field (this
is allowed in $0+1$ dimensions). In that case, the partition function \Z\
would be an infinite series in $\exp(-TM)$ and, correspondingly, $K$ would
contain an infinite tower of representations of $SU(N)$. The first two
terms, however, would be the same as in the fermionic case and we would
obtain the same result.

Ignoring the constant $M$, thus, we again have the same matrix model
Hamiltonian. Unlike, however, the case in [\MP], the physical degrees
of freedom are not merely the diagonal components of $W$, since the commutator
of $W$ with $\dot W$ does not vanish. The commutator in \WWdeq\ is, in fact,
the generator of unitary transformations of $W$ and \WWdeq\ tells us
that the angular degrees of freedom of $W$ are in an ``angular momentum"
state determined by the representation $R$ of $K$. To isolate the
relevant degrees of freedom, let us rewrite $W$ as
$W=U\Lam\Udag$, where $\Lam$ is diagonal. Then the constraint \WWdeq\ leads
to the equation
$$2\UddU-\Lam\UddU\Lamdag-\Lamdag\UddU\Lam=-ig^2LJ,\eqn\UddUeq$$
where $J=\Udag KU$.  Letting $\Om=\UddU$, \UddUeq\ becomes
$$\eqalign{\Om_{ij}\left(2-e^{i(\th_i-\th_j)}-e^{-i(\th_i-\th_j)}\right)&
=-ig^2LJ_{ij},\cr
\Om_{ij}=-ig^2L{J_{ij}\over 4\sin^2(\th_i-\th_j)/2}}\eqn\Omeq$$
where $e^{i\theta_i}$ are the eigenvalues of $\Lam$. Under these substitutions,
the Hamiltonian becomes
$$H={1\over2g^2L}\sum_i {\dot \th}_i^2 +{g^2L\over2}\sum_{i\ne j}
{J_{ij}J_{ji}\over4\sin^2(\th_i-\th_j)/2}.\eqn\newHam$$
Hence this is a theory of fermions on the circle interacting through
two-body forces which depend on $J$. The fermionization is achieved through
the Jacobian factor of the change of variables from $W$ to $(U,\Lam )$.
This introduces the Vandermonde determinant in the wavefunction of the states,
which in the unitary matrix case reads
$$ \Delta = \prod_{i<j} \sin{\theta_i - \theta_j \over 2}.\eqn\VM$$
Each factor in \VM\ is antiperiodic on the circle.
Thus, if $N$ is even the fermions have antiperiodic boundary conditions.
Likewise, if $N$ is odd they have periodic boundary conditions.
This can be understood in terms of transporting a fermion once around the
circle, passing by $N-1$ other fermions along the way
and therefore picking up $N-1$ minus signs.

Notice that $K$ commutes with the Hamiltonian and is a constant of the motion.
The interaction term in \newHam\ on the other hand involves $J=\Udag KU$
rather than $K$. As an operator, $J$ also obeys the $SU(N)$ algebra and
carries the same representation $R$ as $K$. In fact, using \WWdeq, we see that
$K$ acting on physical states generates left-rotations of $U$ and thus
unitary rotations of $W$ while $J$ generates right-rotations of $U$.
It is not, however, a constant of the motion since it does not commute
with the Hamiltonian \newHam.

The form of the Hamiltonian \newHam\ is reminiscent of the Sutherland model
[\Sut], but the coefficients of the interaction terms are particle, color and
time dependent, apparently making a solution impossible. But as it turns out,
there are some residual constraints which will allow us to further
reduce these terms. In fact, \WWdeq, and hence \Omeq, do
not actually contain all of the constraints in the theory. This is because
there are constraints for the diagonal components of $J$, implied by
\gconstraint, which do not show up in the commutation relations. What is
missing is the analog of Gauss' law for the $U(1)$ components of the theory.
In the diagonal basis, Gauss' law \gconstraint\ states that the charges
for each of the diagonal generators must be zero, since space
is compact. This means that the Wilson loop must carry no $U(1)$ charge, but
also that $J$ satisfies on physical states
$$J_{ii}\, |{\rm phys}>=0 ~~~{\rm for~all}~i.\eqn\Jdiag$$
(Alternatively, the diagonal components of $J$ generate diagonal
right-rotations of $U$ which is a redundancy of the parametrization $(U,\Lam )$
since it gives the same $W$, and \Jdiag\ expresses the fact that physical
states are independent of this redundancy.)
One implication of this is that $R$ must be restricted to $SU(N)$
representations with $mN$ boxes in the Young tableau, where $m$ is an integer,
since these are the only representations that have states where all diagonal
charges are zero. This restriction also agrees with the requirement that the
Wilson loop must have zero $Z_N$ charge, else its expectation value vanishes.

To establish the description of the above theory as a system of fermions with
internal degrees of freedom we proceed using two different methods. The first
involves constructing $SU(N)$ representations out of bosonic creation
operators. Let these operators be given by
$\adI_i$, where $i$ is an index that runs from $1$ to $N$ and $I$ is
an index that runs from $1$ to $n$.
The generators can then be expressed as
$$J^a=\sum_I \adI_i\tau^a_{ij}a^I_j.\eqn\gens$$
$SU(N)$ representations are then constructed by acting with the $\adI_i$
on the vacuum state,  so a typical state
looks like
$$a^{\dag 1}_1a^{\dag1}_2.......a^{\dag n}_N|0\rangle.$$
The constraint \Jdiag\ that the diagonal charges all be zero means that the
only allowed states are those where the number of creation operators
for each lower index $i$ is the same for all $i$.  Therefore, each lower
index $i$, which corresponds to a particular fermion, has
an associated state made up of the $n$ different creation operators
$\adI_i$.  This state necessarily transforms as a symmetric representation
of $SU(n)$.  If there are $m$ creation operators per lower index, then
the representation is the $m$-fold symmetric representation.  Hence,
the problem can be thought of as a system of interacting fermions
with internal degrees of freedom which transform in the $m$-fold
symmetric representation of $SU(n)$.

In terms of the creation and annihilation operators, the currents $J_{ij}$
are given by
$$J_{ij}=\sum_I \adI_i a^I_j,\eqn\current$$
hence the symmetrized product $J_{ij}J_{ji}$ is given by
$${1\over2}(J_{ij}J_{ji}+J_{ji}J_{ij})=\sum_{I\ne J}\adI_ia^J_i\adJ_ja^I_j
+\sum_I\adI_i a^I_i \adI_j a^I_j+{1\over2}\sum_I(\adI_ia^I_i+\adI_ja^I_j)
.\eqn\Jsq$$
Clearly, the first sum in \Jsq\ is twice the sum over all nondiagonal
generators
in the product $L^a_iL^a_j$, where $L^a$ are generators of $SU(n)$.  The
next sum contains the diagonal generators as well as an overall constant.
To find their contributions, note that a useful basis for the diagonal
components is
$$M^k={1\over\sqrt{2k(k+1)}}\left(\sum_{l=1}^ka^{\dag l} a^l-ka^{\dag k+1}
a^{k+1}\right).
\eqn\Ldiag$$
Then using the relations
$$\sum_{k=1}^{n-1}{1\over2k(k+1)}={(n-1)\over2n},\eqn\identI$$
and
$$
{-j\over2j(j+1)}+\sum_{k=j+1}^{n-1}{1\over2k(k+1)}=-{1\over2n},\eqn\IdentII$$
one finds that
$$\eqalign{2\sum_{k=1}^{n-1}M^k_iM^k_j&=
{n-1\over n}\sum_{I=1}^n n_i^In_j^I-{1\over n}
\sum_{I\ne J}n_i^In_j^J\cr
&=\sum_In^I_in^I_j-{1\over n}\sum_In^I_i\sum_In^I_j,}\eqn\diagrel$$
where $n_i^I$ is the number operator $\adI_i a^I_i$.
Plugging \diagrel\ into \Jsq, one finds that
$$\eqalign{{1\over2}(J_{ij}J_{ji}+J_{ji}J_{ij})&=m+{m^2/n}+2L^a_iL^a_j\cr
&={2C_{2m}\over(n-1)}+2L^a_iL^a_j,}\eqn\JsqII$$
where $C_{2m}$ is the quadratic casimir for the $m$-fold symmetric
representation of $SU(n)$.  Hence the complete Hamiltonian is
$$H=-{g^2L\over2}\sum_i{\partial^2\over\partial\th_i^2}+{g^2L\over2}
\sum_{i\ne
j}{2C_{2m}/(n-1)+2L^a_iL^a_j\over4\sin^2(\th_i-\th_j)/2}\eqn\Hamfin$$

The particular representation of the Wilson lines determines how many
sets of bosonic operators we should choose.  If the Young tableau
has $n$ rows, then it is necessary to choose at least $n$ sets of
such operators.  Of course we could also build other representations
of $SU(N)$ from these operators, including ones with fewer than $n$ rows.
The thing to notice is that the total $SU(n)$ operator $L^a =\sum_i L^a_i$
commutes with the Hamiltonian. Hence the representation of $SU(N)$ is
determined by the particular irrep of $SU(n)$ chosen for $L^a$.
Since the total operator $L^a$ is in the tensor product of $N$ $m$-fold
symmetric irreps for $SU(n)$, to each irrep in the decomposition of this
product corresponds an irrep of $SU(N)$, found by taking the Young tableau of
the $SU(n)$ irrep and adding to the left enough columns of length $n$ so
as to make the total number of boxes equal to $mN$.

It is instructive to examine a couple of simple examples. Consider, for
instance, representations with $N$ boxes, all in the first or second row.
Thus we choose $n=2$ sets of bosonic operators. Since $m=1$, each fermion
carries the degrees of freedom of a doublet under $SU(2)$.
The only casimir is of course the total spin squared, and since we
have $N$ fermions, the possible total spin states range from $N/2$
down to $1/2$ or $0$, depending on whether $N$ is odd or even.  The
states with maximum total spin correspond to the totally symmetric
representation, and hence has all $N$ boxes in the first row.
The states with total spin ${N \over 2}-l$ are in the representation
with $l$ of its boxes in the second row. The smallest possible spin has the
maximum allowed value in the second row, which is $N/2$ ($(N-1)/2$) for $N$
even (odd). As another example consider the totally symmetric representation
with $mN$ boxes. As was shown in [\APg], this gives rise to the Sutherland
model with integral coefficient. In this case we choose $n=1$ oscillator and
thus there are no internal degrees of freedom. Putting $L=0$ in \JsqII\
we recover the standar coefficient $m(m+1)$ of the Sutherland term.
The connection of \QCD\ and the Sutherland system was also recently
analyzed in [\GN].

The other way to construct states is to use fermionic operators $\bdI_i$.
Hence the generators are
$$J^a=\sum_I \bdI_i\tau^a b^I_j.\eqn\fermgen$$
One can then proceed as before.  In this case, each fermion will
transform in an antisymmetric representation of $SU(n)$.  If the Wilson
line has $n$ columns, then it is necessary to choose at least $n$
sets of $\bdI_i$ operators.  Finally, the Hamiltonian is given as
$$H=-{g^2L\over2}\sum_i{\partial^2\over\partial\th_i^2}+{g^2L\over2}
\sum_{i\ne j}{2\widetilde C_{2m}/(n+1)-2L^a_iL^a_j\over4\sin^2(\th_i-\th_j)/2},
\eqn\Hamfinf$$
where $\widetilde C_{2m}$ is the quadratic casimir for the $m$-fold
antisymmetric representation of $SU(n)$.
The fact that the problem can be expressed either way leads to some
interesting relations between theories, since for a particular representation
we can describe the theory either by the number of rows or by the number
of columns.

It should be stressed that the above theories are fermionic with respect to
total particle exchange (position and internal). This is because there
is yet a discrete remnant of parametrization invariance, namely the one
where $U$ changes by a component that interchanges two eigenvalues and
$\Lam$ changes accordingly. The first transformation exchanges the internal
degrees of freedom of two particles while the second exchanges their
position. Due to the Vandermonde \VM, the wavefunction picks up a minus
sign under these transformations and is thus fermionic.

We now come to one of the main points of this paper. So far we have reduced
\QCD\ with a Wilson loop to a theory of interacting particles but we have
not solved it. By using, however, an alternative reduction of the initial
theory we will explicitly evaluate its spectrum. We will merely consider
the Wilson loop to be spacelike, as we have the freedom to do.
The time evolution then is the one of a free theory, with a particular
weight introduced to initial and final states due to the Wilson loop.
The result for the partition function (as also calculated by Migdal and later
by Rusakov [\Migd,\Rus] using the heat kernel action) is
$$\sum_{R'}\int d U \chi_{R'}(U)\chi_{R'}(\Udag)\chi_R(U)
e^{-g^2LT\CRp},\eqn\MRpi$$
where the sum is over all representations of $U(N)$, $\chi_R(U)$ is the
character for the group element $U$ in representation $R$ and $\CRp$ is the
quadratic casimir for $R'$. The first two characters are the wavefunctions of
the initial and final states while the third is the Wilson loop insertion.

The integral over $U$ of the characters gives an integer
$D(R,R')$ which measures how many times the representation $R'$ is contained in
the the tensor product $R\times R'$. Thus a representation contributes to the
path integral in \MRpi\ only if the tensor product $R\times R'$ contains the
representation $R'$.  This immediately implies that $R$ must have no $U(1)$
charge in order for the path integral to be nonzero.  Moreover,
the number of boxes in the Young tableau that describes this representation
must be a multiple of $N$ in order for there to exist representations $R'$
that satisfy $R'\in R\times R'$, recovering once more this condition.
The casimirs appearing in the exponent of \MRpi, on the other hand,
correspond to the energy eigenvalues of $N$ free nonrelativistic particles.
Thus we conclude that the spectrum of this theory is identical to the
spectrum of a free fermion theory, but with degeneracies $D(R,R')$ determined
by the particular representation $R$.

Again it is instructive to work out explicitly the case where $R$ is the
$m$-fold completely symmetric representation which corresponds to the
Sutherland model. Using standard Young tableau rules, we find that $R'$ can
be contained in the product $R\times R'$ at most once, and that will happen
if the rows $n_i$ of $R'$ satisfy the condition
$$n_i \geq n_{i+1}+m.\eqn\rows$$
Writing $n_i = k_i +(m-1)(N-i)$, we see that $k_i$ must satisfy
$k_{i+1} > k_i$. The second Casimir then is
$$\CRp =\sum_i \half (n_i +N-i)^2 = \sum_i \half k_i^2 +m\sum_{i<j}
|k_i - k_j | + m^2 {N(N-1)(2N-1) \over 12}\eqn\Suth$$
which is indeed the spectrum of the Sutherland model in the ``fermionic"
parametrization of the momenta [\Sut]. It is clear from \rows\ that Sutherland
particles can be thought of as particles whose momenta must be at least
$m+1$ quanta away from each other, putting forth the description of this
model as free fermions with an enhanced exclusion principle.

The above theories then are solvable generalizations of the Sutherland model.
If the number of boxes for the representation $R$ is $N$ then $m=1$ and the
fermions transform in the fundamental representation of $SU(n)$. In this case,
the operator that exchanges the internal quantum numbers is given by
$$\sigma_{ij}={1\over n}+2\tau^a_i\tau^a_j,\eqn\exch$$
hence we can reexpress the Hamiltonians in \Hamfin\ or \Hamfinf\ in
terms of exchange operators. Such a theory has been previously examined
[\Kaw-\MPII]. In particular, in [\MPII] this theory was shown to be
integrable for a generic range of coeficients for the interaction term.
For the higher representations, however, it is not possible to describe
$L^a_iL^a_j$ in terms of an operator that exchanges all of their internal
quantum numbers. But clearly, such a theory must be integrable, at least
for a certain choice of coefficients, since its spectrum can be explicitly
found. It is therefore a challenge to find a proof of integrability
for these particular theories with generic coefficients.

Another point to consider is that,
in general, while the spectra for the Sutherland-like models are known,
the correlation functions have proven to be difficult to compute.
It is possible
that  \QCD\ might provide a useful means for studying this problem.

\ack{The research of J.A.M. was supported in part by D.O.E. grant
DE-AS05-85ER-40518.}
\refout
\end